\documentclass[10pt,twocolumn,letterpaper]{article}

\usepackage[pagenumbers]{cvpr} 

\usepackage[dvipsnames]{xcolor}

\definecolor{cvprblue}{rgb}{0.21,0.49,0.74}
\usepackage[pagebackref,breaklinks,colorlinks,citecolor=cvprblue]{hyperref}

\def\paperID{*****} 
\def\confName{CVPR}
\def\confYear{2024}

\title{Ethics of Generating Synthetic MRI Vocal Tract Views from the Face\\\confName~Responsible Generative AI Workshop}

\author{Muhammad Suhaib Shahid\\
University of Nottingham\\
NG8 1BB, UK \\
{\tt\small Muhammad.Shahid@nottingham.ac.uk}
\and
Gleb E. Yakubov\\
University of Nottingham\\
LE12 5RD, UK \\
\and
Andrew P. French\\
University of Nottingham\\
NG8 1BB, UK\\
}

\begin{document}
\maketitle
\begin{abstract}
Forming oral models capable of understanding the complete dynamics of the oral cavity is vital across research areas such as speech correction, designing foods for the aging population, and dentistry.  Magnetic resonance imaging (MRI) technologies, capable of capturing oral data essential for creating such detailed representations, offer a powerful tool for illustrating articulatory dynamics. However, its real-time application is hindered by expense and expertise requirements. Ever advancing generative AI approaches present themselves as a way to address this barrier by leveraging multi-modal approaches for generating pseudo-MRI views. Nonetheless, this immediately sparks ethical concerns regarding the utilisation of a technology with the capability to produce MRIs from facial observations.

This paper explores the ethical implications of external-to-internal correlation modeling (E2ICM). E2ICM utilises facial movements to infer internal configurations and provides a cost-effective supporting technology for MRI. In this preliminary work, we employ Pix2PixGAN to generate pseudo-MRI views from external articulatory data, demonstrating the feasibility of this approach. Ethical considerations concerning privacy, consent, and potential misuse, which are fundamental to our examination of this innovative methodology, are discussed as a result of this experimentation.
\end{abstract}    
\section{Introduction}
\label{sec:intro}
The ability to model the complete oral cavity holds significant utility across various domains, notably in dentistry, where understanding how dental prosthetics impact speech and mastication is crucial in forming personalised dental devices. However, achieving comprehensive oral cavity modeling presents challenges, including limitations in technique availability, articulator capture capacity, and associated costs. Researchers often face the dilemma of selecting a suitable technique tailored to their specific objectives. Among the available methods, magnetic resonance imaging (MRI) stands out for its capability to provide detailed representations of all articulators, particularly when augmented with real-time functionality, offering insights into dynamic movements. Nonetheless, the practicality of real-time MRI is hindered by its costliness and the need for specialised expertise, rendering it less feasible for routine use.

This raises the question, is it possible to use generative AI approaches to achieve a complete model of the oral cavity, encompassing all articulators in motion, without incurring excessive expenses? This is where the concept of external-to-internal correlation modeling (E2ICM) emerges as a potential solution. By observing the external facial movements, particularly those of visible articulators such as the lips, and jaw, we investigate if we can in any way reconstruct the internal configurations of the oral cavity. This approach leverages the inherent relationship between external facial gestures and internal vocal tract configurations. Such an approach aims to address the cost and complexity concerns associated with MRI and other experimental techniques.  Clearly there are limitations to this approach, but here we consider exploring the feasibility of such a technology, and bring to the fore the ethical questions such an approach might raise.

As advancements in AI-based approaches continue to progress, questions regarding ethical implications become increasingly pertinent. The ability to record or photograph individuals during articulation and mastication, followed by the generation of MRI-like images of the internal oral cavity, raises several ethical considerations and potential concerns regarding privacy, consent, and misuse.

This paper explores the application of generative deep learning models to create pseudo-MRI views of the oral cavity. Specifically, it employs the Pix2PixGAN network to transform external views of a participant during articulation into predicted MRI representations, in a limited speech-reconstruction scenario. For the purpose of demonstrating the feasibility of the proposed approach we briefly evaluate the challenges associated with determining the quality of the generated images. This is followed by a discussion focused on possible ethical dilemmas associated with the use of such generated data.

\section{Background}
\label{sec:formatting}
Real time MRI (RtMRI) of the vocal tract is one of the very few techniques capable of displaying, frame by frame, the movements of all articulators during speech\cite{ramanarayanan_analysis_2018}. The method allows researchers to explore a wide range of applications from articulatory studies, to oral health and food oral processing. Despite the prospects RtMRI presents, there are some underpinning issues that hinder its widespread use. These limitations are a result of the cost and expertise requirements for collecting RtMRI data on an individual subject basis \cite{tiede_contrasts_2000, kochetov_research_2020-1}. One possible generative solution to this is by forming predictive models capable of using external observations of the face to synthesise a representation of the vocal tract MRI view. 

The feasibility of such an approach relies on investigating the interrelationship between the internal vocal tract and external face views. Such research has explored correlations between the two views by linking facial movements, captured via video, with vocal tract dynamics, captured through rtMRI \citep{scholes_inter-relationship_2020}. The main focus is to identify whether there is sufficient mutual information between the forward coronal view of the face and the sagittal MRI to make reconstruction procedures possible. Employing Principal Component Analysis (PCA), Scholes et al. (2020) simplified the data and identified key patterns of change in both modalities. Through this process, they uncovered connections between facial gestures and vocal tract configurations, showcasing the potential for mutual reconstruction between the two modalities. The findings concluded that facial information may hold sufficient data to recover certain vocal tract shapes during speech production. 

While the PCA-based analysis-by-synthesis technique showcases an interrelationship between the two modalities, it comes short of addressing key barriers that prevent the widespread use of MRI. In order to reconstruct an MRI  representation of the vocal tract, a corresponding PCA matrix must accompany each specific external view. However, the PCA representation is derived from the MRI image, and consequently, the MRI data are still necessary each time the representation is created. 

Paving the way to addressing this problem are generative machine learning models capable of performing cross-modality synthesis of unseen MRI configurations when presented with a novel face view for a specific individual \citep{xie_cross-modality_2023}. This technique is commonly used in computer vision and machine learning to create mappings between different visual styles, attributes, or characteristics. It is the process of transforming an input image from one domain into a corresponding output image in another domain, while preserving meaningful content and maintaining consistency between the two domains; it involves changing how an image looks while keeping its underlying meaning intact. In the application of this task, it would involve shifting from a face view to a MRI vocal tract view for any two paired frames; this pairing being key to the approach. 

The Pix2PixGaN framework \cite{isola_image--image_2017} serves as the translation network chosen for this task. The architecture comprises two key components: a generator and a discriminator. The generator works to produce a realistic mapping from the input domain to the desired output, while the discriminator's role is to determine whether an image is real or synthesised. The generator and discriminator train in an adversarial fashion, each trying to optimise ahead of the other. This approach drives the mapping of images from one domain to another in a supervised manner. Once trained, the system has the potential to operate in an autonomous manner and predict internal views based on the outside image or video only. If successful, such approaches can enable generating synthesised views without specialised equipment and a person's consent, which raises important ethical questions and considerations that need to be addressed. 

Existing research has explored the (bio)ethical considerations surrounding the use of Generative Adversarial Networks (GANs) for generating medical images. The integration of AI technologies in healthcare raises complex legal, ethical, and technical challenges. In their work \cite{paladugu_generative_2023}, the authors underscore the necessity for a regulatory framework to ensure the safe integration of generative technologies in medical contexts. A systematic review conducted by \citep{jeong_systematic_2022} examined recent GAN architectures utilised in medical image analysis, revealing imbalances in their capabilities, particularly with smaller datasets. These findings align with the observations of \citep{makhlouf_use_2023}, regarding the imbalanced class distributions often observed in datasets, thereby raising ethical concerns.

\section{Framework and Implementation}
In this study we used a dual-modal dataset used for this study, comprising registered videos captured during speech, this has been previously published\cite{scholes_interrelationship_2020}. Initial data collection involved 13 participants articulating a predetermined set of 10 sentences. Participants underwent two recording sessions: first, speaking the sentences in front of a camera, and second, repeating the same sentences during MRI scans. Subsequently, these video sets were then aligned. Initially, data from 13 participants were collected for the study. However, only data from 11 participants were ultimately included in the published datasets as the study focused on British English speakers. The dataset encompasses videos providing a frontal view of the face alongside sagittal MRI views. For the purposes of this preliminary study, only data from one subject was utilised, as they were the only participant for whom all 10 videos were available across all sentences. Across these 10 videos, a total of 461 frames were available, considering the videos were recorded at a frame rate of 15 frames per second (fps). The shortest video contained 30 frames, while the longest comprised 59 frames.

An implementation of Pix2PixGaN framework was used as the image-to-image translation network. Based on the conditional generative adversarial network (CGaN), the architecture consists of a generator and a discriminator. The generator aims to produce realistic images based on the input, while the discriminator's job is to distinguish between real and generated images. The generator employs a U-Net-inspired encoder-decoder architecture with skip connections. The encoder module is formed of only convolutional layers, omitting dropout. This structure forms the following sequence of layers: C64-C128-C256-C512-C512-C512-C512-C512. The decoder integrates dropout layers with a dropout rate of 0.5 in the first, second, and third layers. The decoder's structure is as follows: CD512-CD512-CD512-C512-C256-C128-C64. This combination establishes a proficient generator capable of producing coherent translations for this dataset. The model was optimised using the Adam optimiser with hyperparameters $\alpha$ = 0.0002, $\beta_1$ = 0.5, $\beta_2$ = 0.999, and $\epsilon$ = 1e-08. The training was done with a batch size of 16, for 200 epochs. Tanh activation function was used.

\section{Results}
The Fréchet Inception Distance (FID) metric and Structural Similarity Index Measure (SSIM) were used to accesses the quality of image examination. FID provides a quantitative measure of similarity between the distribution of generated and authentic images, with lower scores indicating higher quality. SSIM considers the structural information of images, accounting for spatial relationships beyond pixel values. Additionally, a qualitative evaluation was conducted by observing the movements of each articulator frame by frame, drawing conclusions regarding which articulators are constructed most effectively.
The FID score for generated images compared to ground truth is 30.80, though establishing an understanding of what a "good" FID score is when transitioning from RGB to MRI domains remains challenging. To gain some insight, an FID was calculated for various ground truth frames to assess how well the FID performs with real images but of different vocal tract views, yielding a score of 19.75. While FID offers valuable insight into image similarity, it doesn't directly consider the spatial representation of vocal tract structure. Therefore, SSIM might be more suitable for this task. The average SSIM score for all 46 test images was 0.7961, with higher scores indicating better image quality on a scale from -1 to 1. We recognise, and highlight, that interpreting these scores in this application domain is challenging.

As illustrated in Figure 1, the generative models demonstrate some proficiency in generating images with realistic appearances, particularly showcasing discernible movements in the jaw regions. However, upon closer examination, specific articulatory details are challenging to determine. Despite reasonable FID and SSIM scores indicating overall good image similarity, inconsistencies between generated articulators and ground truth are apparent in certain frames. These discrepancies could potentially lead to misleading interpretations in clinical applications, where images resembling plausible MRI scans but with incorrect articulator configurations may pose risks. Moving forward, vocal tract segmentation could serve as a promising avenue for enhancing the clinical relevance in assessing the quality of generated vocal tract views \cite{shahid_research_2024}.

\begin{figure}[t]
  \centering
  \includegraphics[width=1\linewidth]{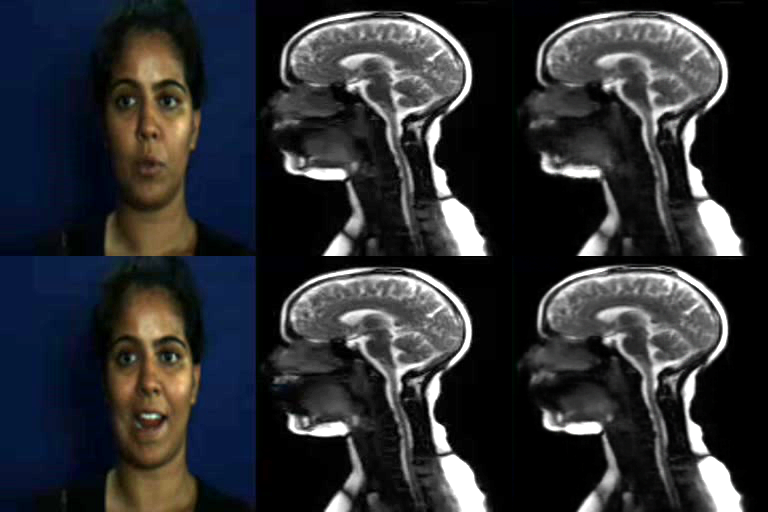}

   \caption{Still frames sample the external view (left),  ground truth MRI frame (middle) and generated frame (right).}
   \label{fig:onecol}
\end{figure}

\section{Discussion of Ethics and Responsible Use}
The ethical dimensions surrounding the potential of such generative medical approaches demand  scrutiny. There are concerns associated with the generation and utilisation of synthetic views, the accuracy and reliability of the data, and potential misuse. 

\subsection{Synthetic dataset enlargement}
The fundamental principle of informed consent and participant autonomy is central. Ethical protocols governing the collection of MRI data are universally stringent, dictating both the type of data collected and its storage practices. Participants are provided with comprehensive information about the study, including potential risks and the intended utilisation of their data, to make informed decisions regarding consent. The integration of generative AI raises logistical and ethical considerations perhaps not originally conceived. In a practical application, MRI data might initially be collected for a limited set of sentences, such as the 10 here. Subsequently, using generative techniques, additional MRI data could be synthesised for sentences that were not originally captured via MRI. When using a trained model, it's possible to employ readily available facial data, especially from public spaces where recording may be allowed by law. However, the ethical dilemma arises: Is it acceptable and responsible to generate new data modalities without obtaining explicit consent? While enlarging datasets using generative AI techniques is not novel and has been applied in various domains, the unique aspect here lies in the translation from an external view to an internal view. Accessing and utilising facial data for such a task, even if publicly available, must be carefully assessed to uphold principles of privacy. 

\subsection{Accuracy of generated images}
Responsible use of generative AI necessitates addressing concerns surrounding the accuracy and integrity of synthesised data. As demonstrated here, methods such as FID are employed to help assess the "quality" of images. However, it is evident from the outset that these methods do not adequately evaluate specific spatial information in the generated images. In other words, the morphology of clinically-relevant structures is not captured well by these metrics.  Often, it is also hard to identify subtle features even directly in the dataset (see Figure 1), so interpreting the quality of synthesised data in this domain is a challenge.

Questions will arise regarding the potential misuse or misinterpretation of inaccurate synthetic information. Poorly performing models could lead to misdiagnosis or misinterpretation. Therefore, it is imperative to remain vigilant and implement rigorous validation procedures to ensure the reliability and accuracy of synthesised data.  More work is needed to develop approaches to assess the usefulness and trustworthiness of generated images.

\subsection{Data storage of generated images}
The stringent protocols governing the storage and anonymisation of MRI data are imperative to safeguard individuals' sensitive health information. However, the creation of \textit{additional} synthetic MRI data, which may still contain identifiable features or morphology, may not always undergo comparable protocol scrutiny as the original data. While large scale MRI datasets could potentially advance medical research and clinical applications, relaxed protocols for synthesised data may compromise privacy and data security. Thus, careful ethical consideration is warranted here for future research. 

Furthermore, there are broader societal implications to consider, particularly regarding the potential impact of synthesised medical views on areas such as identity verification. If such technology were to be deployed in contexts such as security or law enforcement, there could be implications for individuals' rights and freedoms, including the risk of discrimination or misuse of biometric data.

\subsection{Dataset and model biases}

The publicly available dataset used in this study exhibited a bias towards speakers of British English. Though it is understandable from the associated papers that this is likely an attempt to standardise an already small and challenging data set, it nevertheless highlights concerns regarding potential biases in future datasets and the models trained upon them. Certain demographic groups may be favoured in inferences, while for others the model could perform poorly. Likewise, the application of the model would likely be limited to the application studied in the dataset (e.g. speech versus chewing, for example); wrong application would lead to misleading results.

The concern extends beyond  data representation to the broader implications for societal equity and fairness. Though this is a problem  not only relevant to this application, occurrence of such a scenario could also exacerbate existing disparities in access to resources and opportunities, with models not being tailored to regional use. Proactive measures must be implemented to mitigate bias and promote inclusivity in dataset curation and model development. Strategies may include diversifying dataset sources to encompass a broader spectrum of linguistic and cultural backgrounds, and implementing robust validation techniques to identify and mitigate bias in model predictions.

\section{Conclusion}
A demonstration of an exploratory method for generating MRI images of the vocal tract has been presented. Leveraging the Pix2PixGAN architecture, this study demonstrates the application of Generative AI to synthesise previously-unseen vocal tract configurations from external facial views. The quality of the generated images has been evaluated using the Fréchet Inception Distance (FID) metric, alongside the observation of distinct articulator movements.  Results are by no means conclusive at this stage, but certainly raise the question of whether this is a valid line of research in generative AI for future researchers.

Furthermore, an initial discussion regarding the responsible use of generative AI in such applications has been provided. This discussion presents considerations that must be taken into account when employing such techniques. These encompass various aspects, including the enlargement of synthetic datasets, the accuracy of generated images, the storage protocols employed, and the potential biases inherent in both the dataset and the models utilised.
{
    \small
    \bibliographystyle{ieeenat_fullname}
    \bibliography{main}
}


\end{document}